\documentstyle[11pt]{article}
\title{Atmospheric neutrino flux above 1 GeV}
\author{BA-95-49}
\begin{document}
\maketitle
\begin{center}
Vivek Agrawal$^1$, T.K. Gaisser$^1$, Paolo Lipari$^2$ and
                  Todor Stanev$^1$\\[2mm]
$^1$ Bartol Research Institute, University of Delaware, Newark, DE 19716\\
$^2$ Dipartimento di Fisica, Universit\`{a} di Roma, Piazzale Aldo Moro 2,
Rome, Italy\\[2mm]
(submitted to Phys. Rev. D, July, 1995)

\end{center}
\begin{abstract}
In this paper we extend an earlier calculation of the
flux of atmospheric neutrinos to higher energy.  The
earlier calculation of the neutrino flux below 3 GeV has
been used for calculation of the rate of contained neutrino
interactions in deep underground detectors.  The fluxes are needed
up to neutrino energies of 10 TeV to calculate the expected
rate of neutrino-induced muons passing into and through
large, deep detectors.
We compare our results with several other calculations, and
we evaluate the uncertainty in the rate of neutrino-induced muons
due to uncertainties in the neutrino flux.
\end{abstract}

\vspace{3cm}
\noindent
gaisser@bartol.udel.edu

\noindent
lipari@roma1.infn.it

\noindent
stanev@bartol.udel.edu

\newpage
\section{Introduction}
In the past few years several groups have reported new
measurements of $\nu$-induced muons
passing through large, deep-underground
detectors \cite{IMB,Kam,Baksan,Frejus,MACRO,LVD}.
These measurements provide better statistics than the earlier
pioneering measurements at KGF \cite{KGF} and in South Africa
\cite{CWI}, but the results and their interpretation remain
ambiguous, in part because of the intrinsic difficulty of
the measurement but in part because of differences in the
calculations used to interpret the measurements.  In particular,
the relation of these measurements to the
anomalous flavor ratio of contained interactions of atmospheric
neutrinos \cite{Kam1,IMB1} is still controversial.

Interpretation of the flux of neutrino-induced muons depends on
an absolute comparison between a measured rate and a calculated
intensity.  The calculation contains three essential ingredients:
propagation of muons after they are produced in charged-current
interactions of $\nu_\mu$ and $\bar{\nu}_\mu$; the energy spectrum
of muons produced in charged-current interactions of neutrinos
as well as the magnitude of the cross section; and the flux of
atmospheric neutrinos itself.  The first factor in the calculation
is well-understood, and different calculations give similar
results \cite{LKV,LS}.  The uncertainty in the cross section
is discussed in recent papers \cite{FGMS,newIMB,LLS}.  In this paper
we discuss the calculation of the neutrino flux in the energy
range relevant for neutrino-induced muons; that is,
from one GeV up to $10^4$~GeV of neutrino energy.  The calculated
flux of neutrinos and muons given here is essentially an
extension of the low energy calculation \cite{BGS} that has
been used extensively for evaluation of the rate of
contained neutrino interactions.  In this paper we also give
a critical discussion of the sources of uncertainty in the
calculation, as well as a comparison of these results to
those of other calculations.

The neutrino flux depends on the primary cosmic ray spectrum
and on the production of pions and kaons by interactions of
cosmic-ray hadrons in the atmosphere.  Because of the close
relationship between neutrino and muon fluxes, we tabulate
both fluxes.  We also discuss the extent to which measurements
of the muon flux place constraints on the neutrino flux.
We begin in \S2 with a discussion of the primary spectrum
and its uncertainties.  Then in \S3 we discuss the inclusive
cross sections that determine the production of muons and
neutrinos in the atmospheric cosmic ray cascading.  Section 4
contains the neutrino and muon fluxes in tabular form.  In the
conclusion we compare this calculation with others and
summarize our assessment of the uncertainty in the atmospheric
neutrino flux.

\section{Primary spectrum}

The cosmic-ray spectrum incident on the atmosphere consists of
protons and nuclei.  To a first approximation the uncorrelated
spectra of atmospheric secondaries, such as neutrinos and muons,
depend only on the number of incident nucleons in the primary
spectrum as a function of energy-per-nucleon \cite{EGLS}.
We call this the all nucleon spectrum.
The range of neutrino energies important for $\nu$-induced
muons is $1\le E_\nu \le 10^4$~GeV, as shown, for example
in Fig. 2 of Ref. \cite{FGMS}.  The corresponding range of
primary energy-per-nucleon is about a factor of ten higher,
as illustrated in Figure 1.  Approximately 85\% of neutrino-induced
muons come from primary nucleons with energies less than $10^4$~GeV.
(The exact fraction depends somewhat on angle as shown in Fig. 1.)

In this energy range, the all-nucleon spectrum is dominated by
hydrogen and helium, even in the case of extrapolations in
which the spectrum in total energy-per-nucleus (the so-called
``all-particle'' spectrum) is dominated by heavy nuclei at
higher energy.  The relative contribution
of different nuclei to the all-nucleon spectrum  is discussed in
Ref. \cite{GS}, where it is shown, for example,
that hydrogen contributes about 81 (68) \%
of the intensity at $10$ ($10^4$) GeV and helium 72 (70) \% of
the remainder.  Thus the overall uncertainty is dominated by
the uncertainty in the measurements of the spectrum of hydrogen.
The fraction of helium is crucial for determining the charge
ratio of muons and the $\nu/\bar{\nu}$ ratios because this
is the origin of most of the incident neutrons.
Heavy nuclei may become more important around $10^5$~GeV/nucleon,
as noted below.

Figure 2 shows the data summary from Ref. \cite{GS} of the
direct measurements of the primary protons, helium and
heavier nuclei made
with various balloon and satellite experiments.
There are two measurements of the spectrum of hydrogen~\cite{Webber,Seo}
in the 10-100 GeV range that differ in normalization by about 30\%,
an amount which is larger than the statistical error of either experiment.
Up to about $10^4$~GeV, the envelope of the measurements
also covers a range of about $30$\%.  Thus we assign an
uncertainty of $\pm 15$\% to the all-nucleon spectrum in
this entire energy range.

At higher energy the uncertainty of the cosmic ray composition
plays a bigger role.  The data of JACEE~\cite{JACEE1,JACEE2} (shown
by the filled circles in Fig. 2) have
two main features -- a downward bend in the proton spectrum
and an increase of the contribution of all nuclei
heavier than helium.
These two effects tend to compensate each other and leave
the slope of the all nucleon spectrum unchanged up to $10^5$~GeV.
If, however, the bend of the proton spectrum is real but the
flattening of the spectra of the heavy nuclei is not, there will be
a corresponding steepening of the all nucleon spectrum
around $10^4$~GeV.  These two possibilities are indicated
in Fig. 2.  Since primaries with $E>10^4$~GeV/nucleon contribute
only about 15\% of the flux of upward muons, however, uncertainties
at the level of 20 to 30\% above this energy
increase the uncertainty in the
upward muon rate only by a few per cent.
%Thus the uncertainty in the all-nucleon spectrum becomes
%increasingly large above $10^4$~GeV/nucleon--but this
%energy range contributes only about 15\% of the flux
%of upward muons.
%In Fig.~3 we show the fits to the data from Ref. \cite{GS}
%that correspond to these two extrapolations.

For the calculations in this paper we use the all-nucleon
spectrum with the higher extrapolation, which corresponds
to the dash-dot curves in Fig. 2.
This primary spectrum was based originally
on an analysis of the data
summary of Garcia-Munoz and Simpson \cite{Simpson}
done for the calculation of Ref.~\cite{BGS}. It is extended to higher
energy~\cite{GS} by including data from the summary of Swordy~\cite{Swordy}
and other more recent data.
%%%%% \cite{other}.
At energies above $10^4$~GeV/nucleon we use only the measurements
of JACEE~\cite{JACEE1,JACEE2}, shown with filled circles in Fig.~2.
In Fig. 3 we compare the nucleon spectrum used for this
calculation with the other fit to the data of Fig. 2 and
with the spectra used in other calculations of the
neutrino flux at high energy.

The low energy part of the spectrum ($<20$~GeV) is affected
by the geomagnetic field and by modulation by the solar wind,
both of which prevent some fraction of the low energy galactic
cosmic rays from reaching the atmosphere to produce secondaries.
These effects are of greatest importance for the $\sim$~GeV neutrino
flux that is responsible for contained neutrino interactions,
but they also have some importance for neutrino-induced muons,
especially for muons that enter and stop in the detector.
A new evaluation of the geomagnetic cutoffs is the subject
of Ref.~\cite{LS1}.  We tabulate below the reduction in neutrino flux
due to the geomagnetic cutoffs at several detector locations.

\section{Hadron production}

To calculate the atmospheric cascade we use a model of
hadronic interactions called TARGET that is essentially
the same as originally
used in Ref.~\cite{BGS} for calculation of the neutrino
flux below 3 GeV.
It is based on a parameterization of particle
interactions on targets of different mass~\cite{S&O}.  It is
tuned to describe correctly different sets of experimental data
in the tens \cite{Eichten,Allaby,Walker}
to hundreds of GeV \cite{Barton} range of lab energy.
The original version of this model is described in Ref.
\cite{Bangalore}.  The only significant change
since the calculation~\cite{BGS} concerns the description of
of the production of strange hadrons at high energy.  The description of the
production and decay of resonances in the GeV region was also
improved.  The neutrino fluxes below 3 GeV are indistinguishable
from those of Ref. \cite{BGS} in the absence of geomagnetic cutoffs.
A change in the production of kaons in interactions with
energy above $1000$~GeV leads to an increase of the neutrino
flux in the TeV region relative to a preliminary version of this
calculation used in Ref. \cite{FGMS}. On the other hand,
the assumptions of this model about kaon production on nuclear
targets at high energy are at the higher end of the
experimental range and may overestimate kaon production
around 1 TeV and above.  The uncertainty about production of
kaons at high energy is a principal source of uncertainty
in the model which we discuss later.

To get an idea of the uncertainties in the neutrino fluxes
that arise from uncertainties in the description of hadronic
interactions it
is useful to characterize the inclusive cross sections by their
moments, weighted by the shape of the primary spectrum.  For example,
the contribution of pions is approximately proportional to
\begin{equation}
Z_{p\pi^\pm}\;=\;\int_0^1\,x^\gamma{{\rm d}N_{p\pi}\over{\rm d}x}\,
{\rm d}x,
\end{equation}
where ${\rm d}N/{\rm d}x$ is the distribution of charged
pions produced in collisions of protons with
nuclei in the atmosphere, $x=E_\pi/E_p$ and $\gamma$ is the
integral spectral index of the primary cosmic ray spectrum.
The corresponding factors $Z_{pK^\pm}$ (for production of charged
kaons), $Z_{pK^0}$, $Z_{p\pi^0}$, $Z_{pp}$, etc., are defined analogously.

These spectrum weighted moments appear explicitly in analytic approximations
to the uncorrelated particle fluxes in the atmosphere \cite{Book,Lipari}.
Inspection of the analytic approximations for neutrino and
muon fluxes from power-law primary spectra is sufficient to
determine which are the most important sources of uncertainty in different
ranges of primary energy \cite{GLS}.  For vertically incident
leptons with energy below 100 GeV, approximately 95\% of muons and
65\% of $\nu_\mu$
come from pions, whereas above 1000 GeV about 50\% of muons but less than
$\sim 10$\% of $\nu_\mu$ come from decay of pions.  Most of
the remainder come from decay of charged kaons.  (For horizontal
leptons this transition region is shifted to higher energy
\cite{Lipari}, so that
kaons are less important for production of horizontal $\nu_\mu$-induced
muons.)

The relative
importance of kaons for production of neutrinos at high energy
is a consequence of the kinematics of meson decay coupled with
the steep primary spectrum, as explained in Ref. \cite{GLS}.
In the low energy region, the uncertainties in
both the muon and the neutrino
fluxes are dominated by the uncertainty
in pion production, as represented by $Z_{p\pi}$.  At high energy
the dominant source of uncertainty in the neutrino flux is
kaon production, but this is not the case for muons.  As a consequence,
the extent to which measurement of the atmospheric muon spectrum
can be used to normalize the neutrino spectrum is limited.

With this background, we now compare the Z-factors as estimated
from various sets of data and as represented in the TARGET model.
We also quote the Z-factors used for some other
calculations \cite{Minor,VZK} of the
fluxes of high energy neutrinos, where available.
The first six lines of Table~\ref{tab1}~\cite{FHGLS} show estimates
of the Z-factors and related parameters based on three
data sets for proton--proton collisions
at energies from $175$~GeV~\cite{Brenner} to $400$~GeV~\cite{EHS}
on fixed targets and from ISR~\cite{Yen} equivalent to lab energy of
$\sim 1500$~GeV.
The second and third columns
are estimates of the spectrum-weighted moments.
Column 4 is a tabulation of
$R_{K/\pi} = Z_{pK^\pm}/Z_{p\pi^\pm}$.  The factors ${\cal E}(i)$
will be discussed below.

Two estimates of the parameters are given for each of the three data
sets in Table~\ref{tab1}.
For the data of Ref. \cite{Brenner} the first line comes from the
parameterization given in that reference, and second line from
the fit that we performed to the data points.
For Ref. \cite{EHS} the first line is
from direct integration of all data points, and the second excludes
the contribution from x$>$0.6, where the measurements show a strange
feature (possibly due to a contamination of protons in the
samples of positive mesons).
For Ref. \cite{Yen}
the first line is from the analysis of Perkins \cite{Perkins} and
the second from Ref. \cite{Hillas}.

\begin{table}
\caption{Comparison of Z-factors.}
\begin{center}
\begin{tabular}{lcccccc}\hline
Reference      &  $Z_{p\pi^\pm}$ & $Z_{pK^\pm}$ & $R_{K/\pi}$
 & ${\cal E}(\mu)$ & ${\cal E}(\nu_\mu) $ & ${\cal E}(\nu_e)$\\ \hline
 & & & & & &\\
\cite{Brenner} (175 GeV)& 0.074 & 0.0086 & 0.12 & 10.03 & 2.47 & 0.078\\
 &  0.076 & 0.0097 & 0.13 & 10.54 & 2.73 & 0.087 \\
\cite{EHS} (400 GeV)    & 0.079 & 0.0074 & 0.09 & 10.22 & 2.23 & 0.068\\
 & 0.074 &0.0074 & 0.10 & 9.70 & 2.20 & 0.068\\
\cite{Yen} (1500 GeV) & 0.083 & 0.0086 & 0.10 & 10.96 & 2.53 & 0.079\\
 & 0.083 & 0.0100 & 0.12 & 11.34 & 2.85 & 0.090 \\
\cite{Minor} (p--p)  & 0.072 & 0.0094 & 0.13 & 10.04 & 2.64 & 0.085\\
\cite{Minor} (p--air)  & 0.069 & 0.0087 & 0.13 & 9.54 & 2.46 & 0.078\\
\cite{VZK} (p--air) & 0.065 & 0.010 & 0.15 & 9.47 & 2.72 & 0.089 \\
TARGET (1000 GeV)& 0.072 & 0.0105 & 0.15 & 10.34 & 2.89 & 0.094\\ \hline
\end{tabular}
\end{center}
\label{tab1}
\end{table}

In addition to differences among the moments obtained from
data on hydrogen targets,
there is also the uncertainty associated with the relation
between inclusive cross sections on proton targets and
cross sections on light nuclei.  In one case where the
same group took data on both hydrogen and nuclear targets,
the inclusive cross sections at $p_T= 0.3$~GeV/c and
$x = 0.3$ (as near to the peak of the integrand of Eq. 1
as possible with the nuclear target data \cite{Barton})
are similar for the two types of target.
Comparison of $Z$-factors calculated in an event generator
which treats both pp and p-nucleus collisions (SIBYLL)\cite{SIBYLL}
also shows negligible differences between the two.

On the other hand, studies
of the A-dependence of inclusive cross sections on a
variety of nuclear targets (excluding hydrogen) do
show significant variation in the fragmentation
region, with a tendency for the $K/\pi$ ratio to
increase with target mass \cite{Eichten,Marmer,Antreasyan}.
Production of both pions and kaons is enhanced in collisions
on nuclear targets, but kaons may be more enhanced than pions.

Associated production of kaons through $p\rightarrow\Lambda\,K^+$
in the fragmentation region accounts for the large ratio
of $Z_{pK^+}/Z_{pK^-}$ since $K^-$ are produced essentially only through
$K^+/K^-$ pair production in the central region.  The analogous process
for an incident neutron is $n\rightarrow\Lambda\,K^0$.
In TARGET the incident nucleon has an energy-dependent
probability to dissociate into a $\Lambda\,K^+$ pair, which is
assumed to increase from threshold to $4.2$\% at $30$~GeV/c,
$6.5$\% at $300$~GeV/c and asymptotically to $6.8$\%.  By
comparison, the integrated cross section in the forward fragmentation
region for $p+Be\rightarrow\Lambda + X$ at $300$~GeV/c
\cite{Skubic} is $\sim15$~mb, which corresponds to a probability
of $7.3$\% for $\sigma_{p\,Be}=206$~mb.  The probability
for $\Lambda$ production in $pp$ collisions is approximately
5 to 6\% in each hemisphere \cite{Jaeger}.  As a consequence of
the extra channel available for production of positive
kaons by protons, the inclusive cross section is significantly
harder for $K^+$ and for $K^-$ \cite{Bourquin}.

Production of $K^+/K^-$ pairs is determined in TARGET by the
energy-dependent multiplicity of $K^-$ \cite{Antinucci} with an
assumed enhancement of $1.45$ for target ``air'' nuclei ($A=14.5$)
\cite{Marmer}.  The resulting distributions of produced
$\pi^\pm$ and $K^\pm$ agree fairly well with the $19$---$24$~GeV/c
data \cite{Eichten,Allaby}.  In Fig. 4 we compare the distributions
from TARGET for $p$--air interactions at 400 GeV/c with three
sets of $pp$ data (175, 400 and 1500~GeV/c).  As described above,
both pion and kaon production are enhanced at  low $x$ in TARGET, but
kaons more so than pions.  For $x>0.2$ $\pi^+$ distributions
from TARGET (for p-air collisions) are somewhat
below  the $pp$ data and $K^+$ somewhat above.

The $Z$-factors corresponding to TARGET are listed in Table 1,
along with the values from Ref. \cite{Minor,Mitsui}
in which the scheme of Ref. \cite{Tagaki} was used
to relate $pp$ to $p$-air.  The values used by Volkova \cite{Volkova}
to calculate the flux of high energy atmospheric neutrinos are
listed here from Ref. \cite{VZK}.
Considering all the estimates listed in Table 1
(for both $pp$ and $p$-air), the  $Z_{p\pi^\pm}$ cover a range of
approximately $\pm 12$\%.  The corresponding range for
$Z_{pK^\pm}$ is $\pm17$\%.  In the conclusion we use these
numbers to estimate the contribution of the uncertainties
in the input to the neutrino flux calculations to the
uncertainty in the expected flux of neutrino-induced muons.

The last three columns of Table 1 show the influence of these
$Z$--factor sets on the estimates of the uncorrelated fluxes
of muons and neutrinos. ${\cal E}(\mu)$,  ${\cal E}(\nu_\mu)$
and ${\cal E}(\nu_e)$ are the coefficients for the asymptotic
ratio of the vertical lepton flux to the primary cosmic flux
in the relation
$\Phi_{l} = {\cal E}_l \times \Phi_{CR}/E_l$ \cite{Lipari}.
At asymptotically high energy this form gives a good
estimate of the actual muon and neutrino fluxes.
The range of values is $\pm8$\% for muons and
$\pm14$\% for neutrinos.

Finally, there is the question of the energy-dependence of the
$Z$-factors, as shown in Table 2 for TARGET.
For $E<100$~GeV there is energy dependence
(especially for production of strange particles) as the
high-energy, quasi-scaling region is approached.  The
gross energy-dependence
of TARGET above $1000$~GeV is determined by adjusting the
model of Ref. \cite{S&O} so that it also fits measurements
of charged particles produced in
interactions of $20$~TeV protons in lucite \cite{Tasaka}
as described in Ref. \cite{Bangalore}.  The extrapolation reproduces
the rise of the central rapidity plateau that continues
to the energy-range of $\bar{p}p$ colliders.  Since there is
no information on particle production in the fragmentation
region above $1000$~GeV, we make the assumption that the
Z-factors remain constant above this energy.  This is the
difference mentioned at the beginning of this section
that leads to an increase in kaon production compared to
the original version of TARGET.

To estimate the contribution of each Z-factor to the
overall uncertainty in the flux of $\nu_\mu + \bar{\nu}_\mu$
we used the analytic approximations for the neutrino
fluxes from a power law primary spectrum \cite{Book,Lipari}.
The result  is shown as a function of energy in Fig. 5
for estimated maximum uncertainty in each of the various Z-factors
as tabulated in the last column of Table 3.  The other entries
in Table 3 show the relative changes in the flux of
$\nu_\mu+\bar{\nu}_\mu$ due to a fractional change in
the corresponding $Z$-factor.
Most neutrino-induced
upward muons are produced by neutrinos with
$10 \le E_{\nu_\mu} \le 1000$~GeV.  For this reason, the single most
important factor is the uncertainty in $Z_{p\rightarrow K^+}$.

\section{Results of this Calculation}

The results of our
calculation are presented in Tables 4, 5 and 6, for muons,
$\nu_\mu$ and $\nu_e$ respectively. These fluxes correspond
to the cosmic ray spectrum at solar minimum and take no account
of geomagnetic effects. The fluxes are tabulated at energies,
for which other calculations~\cite{Volkova,Mitsui,Butkevich,Honda}
have published their results. More detailed tables are
available on request, which give the fluxes at intervals
of $1/10$ decade of energy.
The fluxes for $E_{lepton}<1000$~GeV are obtained from a straightforward
Monte Carlo calculation, selecting primary neutrons and protons
from the spectrum.  This has the consequence that
statistical fluctuations are noticeable in the tables in the
hundreds of GeV range, especially for $\nu_e$.
At higher energy, where the decay probability
of charged pions and kaons is low, the neutrino flux is obtained by
allowing all mesons to decay and weighting the contributions by the
decay probability.  Thus the statistical accuracy of the
Monte Carlo is better for $E_{lepton}>1000$~GeV.

The fluxes of muon neutrinos presented here are somewhat
greater at high energy than a preliminary version of this calculation
that has been used elsewhere (for example, in Ref. \cite{FGMS})
to calculate the flux of upward, neutrino-induced muons.  This is
a consequence of changing the treatment of
kaon production above $1000$~GeV.  The difference depends
somewhat on angle, but is approximately $+10$\% for
$200<E_\nu<1000$~GeV and $+20$\% for $1 <E_\nu <10$~TeV.
We estimate that this will lead to a $\approx 3$\% increase
in the predicted flux of neutrino-induced
muons with $E_\mu>3$~GeV.

 The quantities presented in Tables 4--6 are intended
for comparisons to other calculations of the atmospheric
neutrino fluxes. For comparisons to experimental data one
should account for the epoch of the solar cycle and the
geomagnetic cutoffs at the location of the detector. For
this purpose we give in Table 7 correction coefficients
for upward going neutrinos at four experimental locations
and for both minimum and maximum solar activity. These corrections
are calculated with a new code for backtracking of cosmic
rays in the IGRF magnetic field  model~\cite{LS1}. The
geomagnetic cutoffs calculated by this code are
quite different from the ones used in Ref.~\cite{BGS}.

\section{Conclusion}
\subsection{Comparison with other calculations}
 Fig.~6a shows our result on the fluxes of vertical muons
of energy between 1 and 1000 GeV compared to the fluxes of
Refs.~\cite{VZK,Butkevich} and to a collection of experimental
data. The slight overestimate of the GeV muon flux in the current
calculation is natural, since no geomagnetic or solar cycle
corrections are applied to the calculation.
The three calculations are in a general agreement, especially
when compared to the $\sim$20\% or more dispersion of the
experimental data. There is a general trend to
overestimate slightly the muon flux between 10 and 100 GeV and
underestimate it at higher energy. The agreement between the
present calculation and Ref.~\cite{Butkevich} is excellent
for muon energy above 100 GeV. This is, however, somewhat
coincidental, since the primary cosmic ray flux of
Ref.~\cite{Butkevich} is significantly different from
the one we use (see Fig. 3).

The second panel in Fig. 6 shows a comparison measurements of
Ref. \cite{Allkofer} of muon fluxes at $0^\circ$ and $75^\circ$.
The good agreement over more than two orders of magnitude
in energy and in flux is a good test of the mechanics of
the calculation, such as the treatment of the atmosphere and
muon decay and energy loss.

 Fig. 7 compares our result for the
angle averaged fluxes of $\nu_\mu + \bar{\nu}_\mu$ with
the results of Refs.~\cite{Volkova,Butkevich,Mitsui}.
Although the general agreement between the four calculations
is not bad, there are significant differences in the
important energy range between 10 and 1000 GeV that
cause differences of order 10 to 15\% in the calculated flux
of upward going neutrino induced GeV muons.

\subsection{Uncertainty in the neutrino flux}
Uncertainties in the calculated neutrino intensity
arise from lack of precise knowledge of the input quantities,
which are the primary spectrum and the inclusive cross section
for production of pions and kaons by hadronic interactions
in the atmosphere.
Because the relative contributions of kaons and
pions to the neutrino flux depends both on energy and
on angle, it is not possible to assign a single estimate
of the uncertainty to the calculation.
We have estimated the primary spectrum uncertainty by a
single overall $\pm15$\%.\footnote{The uncertainties
in the primary spectrum may also change with energy if it
is possible to choose one of the two \cite{Webber,Seo} $10$ to $100$~GeV
measurements over the other.}
If we also were to assign a
similar single uncertainty to the production cross sections,
then we would estimate a overall uncertainty of $\pm21$\%
as in Ref. \cite{FGMS}.

Approximately $70$\% of neutrino-induced muons are produced by
neutrinos with $10 \le E(\nu_\mu) \le 1000$~GeV.  If we focus on
this energy region,
about half the (vertical) neutrinos come from pions and half
from kaons (a larger proportion of kaons above 100 GeV
and a smaller proportion below 100 GeV).
Thus if there is a $\pm12$\% uncertainty in $Z_{p\pi^\pm}$
and and an independent $\pm17$\% uncertainty in $Z_{pK^\pm}$,
then these contribute respectively $\pm6$\% and $\pm8.5$\%.
Combining these uncertainties
with the $\pm15$\% uncertainty
in the primary spectrum as if they were all statistical
errors, we would estimate a $\pm18$\% uncertainty.
For illustration, let us call this nominal result
$1\pm 0.18$.

Another possibility is to use the measured muon spectrum
as a constraint.  As discussed before, this flux depends
only weakly on the properties of kaon production and essentially
measures the product of the primary flux with $Z_{p\pi^\pm}$.
Since the muon on average takes more
energy than the neutrino in $\pi\rightarrow\mu\nu_\mu$,
the relevant range of muon energies is
$30<E_\mu<3000$~GeV.
{}From Fig. 6, we estimate this uncertainty
as $\pm10$\%.  Combining this uncertainty with the
uncertainty in $Z_{pK^\pm}$ as if they were uncorrelated
statistical errors, we find an overall systematic
error of $\pm14$\%.

The errors are not statistical, however.  For example, if
it were determined that the higher set \cite{Webber}
of primary spectrum measurements
below $100$~GeV were correct and the spectral index could
be determined with sufficient precision to extrapolate to higher energy,
we would shift the central value upward by $12$\%
and assign a smaller error $\sim\pm10$\% to the spectrum.
Relative to the nominal estimate above,
our new estimate would be $1.12\pm0.16$, assuming the
other uncertainties remain unchanged.

Finally,
by a similar argument, if we shift the estimate of $Z_{pK^\pm}$
downward by $20$\%, for example, then (always averaging over the
energy range relevant for vertical  upward neutrino-induced
muons) the central value would drop  from unity to $0.9$.
If the uncertainties remain as initially assumed, the new
estimate would be $0.9\pm0.16$.

{\bf Acknowledgements}. We are grateful for helpful discussions
with D. Michael and S. Mikheyev.  This work is supported in
part by the U.S. Department of Energy (TKG and TS) and by the
INFN (PL).

\newpage
\begin{center}
FIGURES
\end{center}
\noindent
FIG. 1.
Primary cosmic ray nucleon energy contribution to the upward going
neutrino induced muon flux ($E_\mu>$1 GeV). The solid line is for
$\cos(\theta)$=--1 and the dashed is for $\cos(\theta)$=--0.15.

\vspace{.5cm}
\noindent
FIG. 2.
 Direct data on the spectra of different comic ray nuclei.
 The data for H and He are from: open circles -- Ref.~\cite{Webber},
inverted triangles -- Ref.~\cite{Seo}, triangles -- Ref.~\cite{Ryan},
filled squares -- Ref.~\cite{Ivanenko}, filled circles --
Ref.~\cite{JACEE1,JACEE2},
crosses -- Ref.~\cite{Kawamura}, hexagons -- Ref.~\cite{Zatsepin} and
open squares -- Ref.~\cite{RICH1}. The data for heavier nuclei are from:
open circles -- Ref.~\cite{HEAO}, triangles -- Ref.~\cite{Simon},
open squares -- Ref.~\cite{RICH2}, filled squares -- Ref.~\cite{CRN},
crosses -- Ref.~\cite{Ichimura} and filled circles -- Ref.~\cite{JACEE2}.
The lines represent the two fits discussed in the text: 1) (solid line)
steepening H and all nucleon spectrum; 2) (dash-dot) a gradual bending of
the H spectrum which is compensated by flattening of the spectra of all
heavier nuclei.

\vspace{.5cm}
\noindent
FIG. 3.
The all nucleon spectra derived from the fits from Fig.~2
(same line coding) compared with spectra used in other neutrino flux
calculations: short dash -- Ref.~\cite{Volkova}, long dash --
Ref.~\cite{Butkevich}, dots -- Ref.~\cite{Mitsui} and  short--long dashes
-- Ref.~\cite{Honda}.  The two data points are from the JACEE data
set~\cite{JACEE1,JACEE2}.

\vspace{.5cm}
\noindent
FIG. 4.
Inclusive cross sections integrated over transverse momentum
derived from experimental
data on $pp$ ($\bar{p}p$) interactions at incident energy of
175 GeV (triangles)~\cite{Brenner}, 400 GeV (circles)~\cite{EHS}  and
an equivalent Lab energy of 1500 GeV (ISR, squares)~\cite{Yen}.
Distributions from the event generator TARGET for p-air collisions
at 400 GeV are shown for comparison.

\vspace{.5cm}
\noindent
FIG. 5
Ratios of the $\nu_\mu + \bar{\nu}_\mu$ flux calculated with different
$Z$--factors increased by their maximum experimental uncertainty
to the flux calculated with the central $Z$--factor values.
The ratios are symmetric, i.e. inverted when the calculation is done
with the lowest allowed $Z$--factor values.

\vspace{.5cm}
\noindent
FIG. 6.
a). Vertical muon fluxes calculated in Refs.~\cite{VZK} (dotted lines
with high and low normalization) and~\cite{Butkevich} are compared with
the current calculation (solid line) and experimental data. The data
points are from Refs.~\cite{Allkofer} (squares), \cite{Rastin} (hexagons),
\cite{Ayre}(diamonds) and \cite{Ivanenmu} (pentagons).
b). Muon flux at $\theta=0^\circ$ (diamonds) \cite{Allkofer} and
$\theta=75^\circ$ (squares) \cite{Jokisch} compared to the current
calculation.

\vspace{.5cm}
\noindent
FIG. 7.
a). Angle average fluxes of $\nu_\mu + \bar{\nu}_\mu$ calculated by
Volkova~\cite{Volkova} (short dash), Butkevich {\it et al.}~\cite{Butkevich}
(long dash), Mitsui {\it et al.}~\cite{Mitsui} (dots) and Honda {\it et al.}
\cite{Honda} (short-long dash) compared to the this calculation.
b). Ratio of the fluxes from Fig.~7a to the ones presented here.
The line coding is the same.

\newpage
\noindent
           Table~2. $Z$ factors from the TARGET event generators for
      proton interactions on air nuclei for\\ $\gamma$ = 1.7\\[5mm]
\begin{small}
\begin{center}
\begin{tabular}{|l||c|c|c|c|c|c|c|c|}\hline
E$_p$  & \multicolumn{8}{c}{Secondary particles} \\ \hline
 (GeV) & $p$ & $\pi^+$ & $\pi^-$ & $K^+$ & $K^-$ & $\pi^0$ & $ K^0 $ & $n$
\\ \hline \hline
 &  &  &  &  &  & & & \\
  3      &0.4000&0.0239&0.0154&0.0001&0.0000&0.0200&0.0001&0.1013 \\
  10     &0.2742&0.0529&0.0365&0.0014&0.0000&0.0447&0.0008&0.0352 \\
  10$^2$ &0.2681&0.0456&0.0329&0.0072&0.0029&0.0390&0.0082&0.0335 \\
  10$^3$ &0.2732&0.0414&0.0302&0.0079&0.0027&0.0356&0.0071&0.0372 \\
  10$^4$ &0.2710&0.0405&0.0301&0.0076&0.0026&0.0346&0.0070&0.0369 \\
  10$^5$ &0.2696&0.0409&0.0292&0.0076&0.0025&0.0346&0.0068&0.0372 \\
 &  &  &  &  &  & & & \\ \hline
\end{tabular}
\end{center}
\end{small}
%\end{center}
\newpage
\noindent
 Table~3. Effect of different $Z$ factors on the flux
 of muon neutrinos and antineutrinos. Only $Z$ factors
 affecting the flux by more than 2\% are included. The
 last column gives the maximum relative uncertainty
 deduced from experimental data and used for the
 estimate shown in Fig. 5.
\\[5mm]
\begin{center}
\begin{tabular}{|l||c|c|c|c|}\hline
 $Z$ & \multicolumn{3}{c|}{$ \partial  F (\nu_\mu + \bar{\nu}_\mu) /
\partial \ln{Z} $} & $\Delta Z/Z$  \\  \cline{2-4}
\ \ \ \ \ $E_\nu$, GeV& 10. & 100. & 1000. & \\ \hline \hline
 &  &  &  &  \\
 $pp$         & 0.338 & 0.307 & 0.310 & 0.14 \\
 $pn$         & 0.046 & 0.040 & 0.030 & 0.68 \\
 $p\pi^+$     & 0.482 & 0.359 & 0.186 & 0.12 \\
 $p\pi^-$     & 0.276 & 0.222 & 0.146 & 0.12 \\
 $pK^+$       & 0.163 & 0.273 & 0.436 & 0.25 \\
 $pK^-$       & 0.071 & 0.118 & 0.188 & 0.25 \\
 $\pi^+\pi^+$ & 0.019 & 0.047 & 0.048 & 0.19 \\
 $\pi^+K^+$   & 0.003 & 0.037 & 0.045 & 0.29 \\
 $\pi^+K^-$   & 0.002 & 0.020 & 0.024 & 0.25 \\
 &  &  &  &  \\
\hline
\end{tabular}
\end{center}
\newpage
\noindent
 Table~4. Fluxes of atmospheric muons as a function
 of the zenith angle. The values shown are d$N_\mu$/d($\ln E_\mu$) in units
of cm$^{-2}$s$^{-1}$srad$^{-1}$.
\\[5mm]
\begin{small}
\begin{center}
\begin{tabular}{|r||c|c|c|c|c|c|}\hline
cos$\theta$& 1.0 & 0.75 & 0.50 & 0.25 & 0.15 & 0.05 \\ \hline
$E_\nu$, GeV & \multicolumn{6}{c|} {} \\ \hline \hline
 &  &  &  &  &  &  \\
 1.&4.03E-03&1.63E-03&3.78E-04&3.15E-05&2.49E-06&7.16E-07\\
 2.&4.11E-03&1.98E-03&6.59E-04&6.41E-05&1.27E-05&1.20E-06\\
 3.&3.59E-03&2.03E-03&7.47E-04&9.46E-05&2.38E-05&1.68E-06\\
 5.&2.61E-03&1.68E-03&7.87E-04&1.43E-04&3.44E-05&2.92E-06\\
 10.&1.33E-03&1.02E-03&6.19E-04&1.76E-04&6.27E-05&5.47E-06\\
 20.&5.29E-04&4.63E-04&3.51E-04&1.55E-04&7.02E-05&9.97E-06\\
 30.&2.80E-04&2.60E-04&2.18E-04&1.20E-04&6.37E-05&1.24E-05\\
 50.&1.15E-04&1.15E-04&1.07E-04&7.46E-05&4.97E-05&1.38E-05\\
 100.&2.94E-05&3.22E-05&3.39E-05&3.09E-05&2.50E-05&1.11E-05\\
 200.&6.45E-06&7.51E-06&8.90E-06&9.97E-06&9.48E-06&6.27E-06\\
 300.&2.54E-06&3.04E-06&3.73E-06&4.73E-06&4.75E-06&3.76E-06\\
 500.&7.33E-07&9.17E-07&1.19E-06&1.70E-06&1.77E-06&1.69E-06\\
 1000.&1.30E-07&1.66E-07&2.29E-07&3.69E-07&3.99E-07&4.58E-07\\
 2000.&2.24E-08&2.92E-08&4.09E-08&7.02E-08&7.85E-08&1.01E-07\\
 3000.&7.66E-09&9.82E-09&1.40E-08&2.63E-08&2.91E-08&3.87E-08\\
 5000.&1.96E-09&2.89E-09&3.84E-09&7.32E-09&7.87E-09&1.10E-08\\
 &  &  &  &  &  &  \\
\hline
\end{tabular}  \end{center}
\end{small}
\newpage
\noindent
 Table~5. Fluxes of atmospheric $\nu_\mu$ and $\bar{\nu}_\mu$ as a function
 of the zenith angle. The values shown are d$N_\nu$/d($\ln E_\nu$) in units
of cm$^{-2}$s$^{-1}$srad$^{-1}$.
\\[5mm]
\begin{tiny}
\begin{center}
\begin{tabular}{|r||c|c||c|c||c|c||c|c||c|c||c|c|}\hline
cos$\theta$&\multicolumn{2}{c||}{1.00}&\multicolumn{2}{c||}{0.75}
&\multicolumn{2}{c||}{0.50}&\multicolumn{2}{c||}{0.25}
&\multicolumn{2}{c||}{0.15}&\multicolumn{2}{c|}{0.05}\\ \hline
$E_\nu$, GeV &$\nu_\mu$&$\bar{\nu}_\mu$&$\nu_\mu$&$\bar{\nu}_\mu$
&$\nu_\mu$&$\bar{\nu}_\mu$&$\nu_\mu$&$\bar{\nu}_\mu$
&$\nu_\mu$&$\bar{\nu}_\mu$&$\nu_\mu$&$\bar{\nu}_\mu$ \\ \hline \hline
 &  &  &  &  &  &  &  &  &  &  &  & \\

1.&1.88E-02&1.74E-02&1.99E-02&1.90E-02&2.09E-02&2.08E-02&2.24E-02&2.29E-02&2.35E-02&2.40E-02&2.40E-02&2.38E-02\\

2.&5.38E-03&4.97E-03&5.79E-03&5.53E-03&6.31E-03&6.16E-03&7.01E-03&7.08E-03&7.45E-03&7.61E-03&7.78E-03&7.77E-03\\

3.&2.47E-03&2.23E-03&2.74E-03&2.55E-03&3.03E-03&2.89E-03&3.47E-03&3.38E-03&3.65E-03&3.68E-03&3.83E-03&3.89E-03\\

5.&9.63E-04&8.36E-04&1.07E-03&9.51E-04&1.20E-03&1.07E-03&1.41E-03&1.34E-03&1.50E-03&1.47E-03&1.61E-03&1.63E-03\\

10.&2.77E-04&2.21E-04&2.89E-04&2.46E-04&3.31E-04&2.78E-04&3.89E-04&3.55E-04&4.22E-04&3.94E-04&4.74E-04&4.59E-04\\

20.&7.19E-05&5.46E-05&7.65E-05&6.16E-05&8.72E-05&7.18E-05&1.06E-04&9.03E-05&1.15E-04&1.01E-04&1.33E-04&1.23E-04\\

30.&3.23E-05&2.34E-05&3.58E-05&2.59E-05&4.01E-05&3.07E-05&4.74E-05&3.95E-05&5.34E-05&4.53E-05&6.10E-05&5.58E-05\\

50.&1.16E-05&7.99E-06&1.28E-05&9.07E-06&1.45E-05&1.08E-05&1.77E-05&1.38E-05&1.98E-05&1.60E-05&2.34E-05&1.98E-05\\

100.&2.80E-06&1.73E-06&3.11E-06&1.99E-06&3.59E-06&2.39E-06&4.51E-06&3.20E-06&5.00E-06&3.74E-06&6.00E-06&4.80E-06\\

200.&6.71E-07&3.67E-07&7.37E-07&4.26E-07&8.62E-07&5.28E-07&1.10E-06&7.04E-07&1.23E-06&8.29E-07&1.49E-06&1.09E-06\\

300.&2.69E-07&1.43E-07&3.21E-07&1.75E-07&3.71E-07&2.14E-07&4.82E-07&3.02E-07&5.34E-07&3.40E-07&6.49E-07&4.45E-07\\

500.&8.95E-08&4.40E-08&1.02E-07&5.23E-08&1.25E-07&6.64E-08&1.67E-07&9.93E-08&1.82E-07&1.09E-07&2.18E-07&1.43E-07\\

1000.&1.82E-08&8.37E-09&2.18E-08&1.04E-08&2.72E-08&1.34E-08&3.71E-08&1.91E-08&4.26E-08&2.29E-08&5.07E-08&3.01E-08\\

2000.&3.33E-09&1.50E-09&4.10E-09&1.87E-09&5.53E-09&2.58E-09&7.93E-09&3.88E-09&9.39E-09&4.72E-09&1.14E-08&6.01E-09\\

3000.&1.21E-09&5.33E-10&1.53E-09&6.73E-10&2.08E-09&9.59E-10&3.12E-09&1.44E-09&3.77E-09&1.72E-09&4.75E-09&2.45E-09\\

5000.&3.33E-10&1.41E-10&4.28E-10&1.80E-10&5.98E-10&2.60E-10&9.37E-10&4.20E-10&1.14E-09&5.03E-10&1.50E-09&6.91E-10\\

10000.&5.27E-11&2.21E-11&6.89E-11&2.86E-11&1.01E-10&4.32E-11&1.71E-10&7.07E-11&2.08E-10&8.82E-11&2.91E-10&1.25E-10\\
  &  &  &  &  &  &  &  &  &  &  &  & \\
\hline
\end{tabular}  \end{center}
\end{tiny}
\newpage
\noindent
 Table~6. Fluxes of atmospheric $\nu_e$ and $\bar{\nu}_e$ as a function
 of the zenith angle. The values shown are d$N_\nu$/d($\ln E_\nu$) in units
of cm$^{-2}$s$^{-1}$srad$^{-1}$.
\\[5mm]
\begin{tiny}
\begin{center}
\begin{tabular}{|r||c|c||c|c||c|c||c|c||c|c||c|c|}\hline
cos$\theta$&\multicolumn{2}{c||}{1.00}&\multicolumn{2}{c||}{0.75}
&\multicolumn{2}{c||}{0.50}&\multicolumn{2}{c||}{0.25}
&\multicolumn{2}{c||}{0.15}&\multicolumn{2}{c|}{0.05}\\ \hline
$E_\nu$, GeV &$\nu_e$&$\bar{\nu}_e$&$\nu_e$&$\bar{\nu}_e$
&$\nu_e$&$\bar{\nu}_e$&$\nu_e$&$\bar{\nu}_e$
&$\nu_e$&$\bar{\nu}_e$&$\nu_e$&$\bar{\nu}_e$ \\ \hline \hline
 &  &  &  &  &  &  &  &  &  &  &  & \\

1.&8.23E-03&6.51E-03&9.75E-03&7.24E-03&1.15E-02&8.85E-03&1.36E-02&1.02E-02&1.40E-02&1.15E-02&1.48E-02&1.17E-02\\

2.&2.03E-03&1.56E-03&2.54E-03&2.00E-03&3.16E-03&2.46E-03&4.09E-03&3.26E-03&4.58E-03&3.55E-03&4.89E-03&3.83E-03\\

3.&7.87E-04&6.66E-04&1.04E-03&8.17E-04&1.36E-03&1.11E-03&1.90E-03&1.48E-03&2.16E-03&1.68E-03&2.38E-03&1.90E-03\\

5.&2.50E-04&2.02E-04&3.09E-04&2.58E-04&4.35E-04&3.52E-04&6.72E-04&5.42E-04&7.93E-04&6.50E-04&9.47E-04&7.54E-04\\

10.&4.15E-05&3.68E-05&5.99E-05&4.83E-05&8.60E-05&7.18E-05&1.46E-04&1.21E-04&1.88E-04&1.56E-04&2.47E-04&2.02E-04\\

20.&7.77E-06&6.47E-06&1.01E-05&8.39E-06&1.57E-05&1.30E-05&2.89E-05&2.35E-05&3.98E-05&3.22E-05&5.84E-05&4.70E-05\\

30.&2.82E-06&2.35E-06&3.66E-06&3.03E-06&5.76E-06&4.61E-06&1.10E-05&9.05E-06&1.52E-05&1.27E-05&2.53E-05&1.99E-05\\

50.&7.44E-07&6.25E-07&1.05E-06&7.80E-07&1.51E-06&1.22E-06&2.97E-06&2.43E-06&4.26E-06&3.65E-06&7.28E-06&6.25E-06\\

100.&1.48E-07&1.16E-07&1.77E-07&1.52E-07&2.67E-07&1.96E-07&4.89E-07&4.18E-07&7.43E-07&6.30E-07&1.40E-06&1.12E-06\\

200.&2.46E-08&2.37E-08&3.59E-08&2.55E-08&5.01E-08&3.87E-08&8.48E-08&6.79E-08&1.16E-07&9.16E-08&2.43E-07&1.87E-07\\

300.&1.14E-08&7.41E-09&1.25E-08&1.03E-08&2.01E-08&1.22E-08&3.05E-08&2.49E-08&4.19E-08&3.34E-08&7.62E-08&6.28E-08\\

500.&3.29E-09&2.15E-09&3.83E-09&2.63E-09&5.26E-09&3.90E-09&8.42E-09&6.94E-09&1.11E-08&8.57E-09&2.00E-08&1.62E-08\\

1000.&5.86E-10&3.67E-10&7.46E-10&4.58E-10&9.70E-10&6.29E-10&1.52E-09&9.78E-10&1.98E-09&1.47E-09&3.30E-09&2.61E-09\\

2000.&1.00E-10&6.23E-11&1.26E-10&7.80E-11&1.75E-10&1.14E-10&2.68E-10&1.78E-10&3.86E-10&2.42E-10&5.20E-10&4.02E-10\\

3000.&3.51E-11&2.15E-11&4.50E-11&2.71E-11&6.42E-11&4.06E-11&1.01E-10&6.45E-11&1.28E-10&8.35E-11&2.01E-10&1.40E-10\\

5000.&9.24E-12&5.45E-12&1.18E-11&7.01E-12&1.73E-11&1.07E-11&2.86E-11&1.79E-11&3.55E-11&2.25E-11&5.41E-11&3.40E-11\\

10000.&1.47E-12&8.46E-13&1.90E-12&1.13E-12&2.79E-12&1.67E-12&4.84E-12&2.93E-12&6.05E-12&3.63E-12&8.76E-12&5.38E-12\\
 &  &  &  &  &  &  &  &  &  &  &  & \\
\hline
\end{tabular}  \end{center}
\end{tiny}
\newpage
\noindent
 Table~7. Geomagnetic corrections for upward going muon neutrinos
 plus antineutrinos for several experimental locations.\\[5mm]
\begin{tiny}
\begin{center}
\begin{tabular}{|l||r||c|c|c|c|c|c||c|c|c|c|c|c|}\hline
\multicolumn{2}{|l||}{Location}&\multicolumn{6}{c||}{$\odot_{max}$}&
                 \multicolumn{6}{c|}{$\odot_{min}$}\\ \cline{3-14}
 \multicolumn{2}{|r||}{$\cos \theta$}
 & -1.00 & -0.75 & -0.50 & -0.25 & -0.15 & -0.05
 & -1.00 & -0.75 & -0.50 & -0.25 & -0.15 & -0.05 \\ \cline {3-14}
 \multicolumn{2}{|r||}{$E_\nu$, GeV}& \multicolumn{12}{c|}{} \\ \hline \hline
 &  &  &  &  &  &  &  &  &  &  &  &  & \\
 Kamioka &
    1.0&0.75&0.70&0.69&0.64&0.60&0.56 & 0.75&0.70&0.69&0.64&0.61&0.57\\
 &  2.0&0.92&0.88&0.87&0.83&0.78&0.75 & 0.92&0.88&0.88&0.84&0.79&0.75\\
 &  3.0&0.99&0.96&0.95&0.90&0.88&0.85 & 1.00&0.96&0.96&0.90&0.88&0.85\\
 &  5.0&1.00&0.98&1.00&0.94&0.94&0.93 & 1.00&0.98&1.00&0.94&0.94&0.93\\
 &  &  &  &  &  &  &  &  &  &  &  &  & \\
 IMB &
    1.0&0.89&0.90&0.91&0.91&0.87&0.78 & 0.98&0.99&1.00&1.00&0.94&0.84\\
 &  2.0&0.96&0.96&0.97&0.97&0.95&0.90 & 0.98&0.99&1.00&1.00&0.97&0.91\\
 &  3.0&1.00&0.98&1.00&0.98&0.99&0.96 & 1.00&0.99&1.00&1.00&0.99&0.96\\
 &  5.0&1.00&0.99&1.00&0.99&0.99&0.99 & 1.00&1.00&1.00&1.00&0.99&0.99\\
 &  &  &  &  &  &  &  &  &  &  &  &  & \\
 GS/ &
    1.0&0.86&0.87&0.87&0.82&0.75&0.68 & 0.92&0.93&0.92&0.86&0.79&0.72\\
 Frejus &
    2.0&0.96&0.95&0.96&0.95&0.88&0.83 & 0.98&0.97&0.98&0.96&0.90&0.84\\
 &  3.0&1.00&0.98&1.00&0.97&0.95&0.91 & 1.00&1.00&1.00&0.97&0.96&0.91\\
 &  5.0&1.00&0.99&1.00&0.98&0.97&0.96 & 1.00&1.00&1.00&0.99&0.98&0.97\\
 &  &  &  &  &  &  &  &  &  &  &  &  & \\
 SNO/ &
    1.0&0.89&0.90&0.90&0.92&0.89&0.83 & 0.98&0.99&1.00&1.00&0.97&0.89\\
 Soudan &
    2.0&0.96&0.96&0.97&0.98&0.96&0.93 & 0.98&0.99&1.00&1.00&0.98&0.94\\
 &  3.0&1.00&0.99&1.00&0.99&0.99&0.98 & 1.00&1.00&1.00&1.00&1.00&0.98\\
 &  5.0&1.00&1.00&1.00&1.00&0.99&0.99 & 1.00&1.00&1.00&1.00&1.00&0.99\\
 &  &  &  &  &  &  &  &  &  &  &  &  & \\
\hline
\end{tabular}  \end{center}
\end{tiny}

\end{document}